\documentclass[]{raa}            

\usepackage{amsmath,amssymb,amsxtra,amsfonts}
\usepackage{graphicx,times}
\usepackage{epsfig}
\usepackage{natbib}
\usepackage{txfonts}
\usepackage{footmisc}

\begin{document}

   \title{Four New Observational $H(z)$ Data From Luminous Red Galaxies of Sloan Digital Sky Survey Data Release Seven}

\author{Cong Zhang\inst{1}, Han Zhang\inst{1},  Shuo Yuan\inst{1}, Siqi Liu \inst{1}, Tong-Jie Zhang \inst{1} and Yan-Chun Sun\inst{1}}


\institute{ Department of Astronomy, Beijing Normal University, Beijing 100875, P. R. China;
    {\it tjzhang@bnu.edu.cn}\\
       }


\abstract{By adopting the differential age method, we utilize selected 17832 luminous red galaxies (LRGs) from Sloan Digital Sky Survey Data Release Seven (SDSS DR7) covering redshift $0<z<0.4$ to measure Hubble parameters. Using a full spectrum fitting package \texttt{UlySS}, these spectra are reduced with single stellar population (SSP) models and optimal age information of our selected sample are derived. With the decreasing age-redshift relation, four new observational $H(z)$ data (OHD) points are obtained, which are $H(z)=69.0\pm19.6$ km\ s$^{-1}$ Mpc$^{-1}$ at $z=0.07$, $H(z)=68.6\pm26.2$ km\ s$^{-1}$ Mpc$^{-1}$ at $z=0.12$, $H(z)$=$72.9\pm29.6$ km\ s$^{-1}$ Mpc$^{-1}$ at $z=0.2$ and $H(z)$=$88.8\pm36.6$ km\ s$^{-1}$ Mpc$^{-1}$ at $z=0.28$, respectively. Combined with other 21 available OHD data points, a performance of constraint on both flat and non-flat $\Lambda$CDM model is presented.
}

   \authorrunning{Zhang et al. }            
   \titlerunning{Hubble Parameter}  
   \maketitle

\section{Introduction}
\label{sec:one}
  A variety of cosmological observations are used for a better understanding of the expansion of the Universe quantitatively, for example the mapping of the cosmic microwave background (CMB) anisotropies \citep{2007ApJS..170..377S,2011ApJS..192...18K}, the measurement of baryon acoustic oscillation (BAO) peaks \citep{2005ApJ...633..560E,2010MNRAS.401.2148P}, measurements of `standard candles' such as the redshift-distance relationship of type Ia supernovae (SNIa) \citep[SNIa;][]{1998AJ....116.1009R,2009ApJ...700.1097H} and gamma-ray bursts (GRBs) \citep[GRBs;][]{2004ApJ...613L..13G,2008ApJ...680...92L} . Hubble parameter $H(z)$, which is defined as : $H(z)=\dot{a}/a$, where $a$ denotes the cosmic scale factor and $\dot{a}$ is its rate of change with respect to the cosmic time, is directly related to the expansion history of the Universe. The method based on the observational $H(z)$ data (OHD) has been used to test cosmological models
 \citep[e.g.,][]{2007MPLA...22...41Y,2011PhLB..703..406C}. Besides parameters constraints, OHD can also be used as an auxiliary model selection criterion \citep{2009JCAP...06..036L}.

 In practice, the Hubble parameter $H(z)$ is usually evaluated as a function of the redshift
$z$, with $a(t)/a(t_0)=1/(1+z)$, where $t_{0}$ is the current cosmic time:
\begin{equation}
\label{eq:OHDe}
H(z)=-\frac{1}{1 + z}\frac{dz}{dt}.
\end{equation}
$z$ is the cosmological redshift and $t$ is the age of the Universe when the observed photon is emitted.
Derivative of redshift with respect to cosmic time, $dz/dt$ has a direct determination on $H(z)$.
$H(z)$ has been measured through the differential method according to Eq. (\ref{eq:OHDe}), which was first put forward by \citet{2002ApJ...573...37J}. This differential method has been demonstrated in \cite{2003ApJ...593..622J}.
However, it may be difficult to select galaxies as `cosmic chronometers' and determine the accurate age of a galaxy considering stars in a galaxy born continuously, and a young stellar population may dominate their emission spectra \citep{2010AdAst2010E..81Z}. Luminous red galaxies (LRGs) which have photometric properties consistent with an old, passively evolving stellar population \citep{2006MNRAS.373..349R} are regarded as a good candidate of this `cosmic chronometers' \citep{2010MNRAS.406.2569C}.

We employ \texttt{ULySS}\footnote{\texttt{ULySS} is available at: http://ulyss.univ-lyon1.fr/}, an available package on-line to fit a full-length spectrum. As the relatively homogeneous stellar component of LRGs, we use single stellar population (SSP) fitting and gain age information, which will be discussed in detail in Section \ref{sec:third}. With the age-redshift relation, four OHD points are deduced accordingly.

There are already 21 OHD points got from both differential age method \citep{2003ApJ...593..622J, 2005PhRvD..71l3001S,2010JCAP...02..008S, 2012JCAP...08..006M} and baryon acoustic oscillation (BAO) method \citep{2009MNRAS.399.1663G}
. Currently the number of OHD points are still scarce compared with SNIa luminosity distance data. The potential power of OHD in constraining cosmological parameters is explored in \cite{2011ApJ...730...74M} in detail. It has achieved a conclusion that the constraining power of OHD  can be as strong as that of SNIa when its quantity reaches a certain value which, depending on the error model used in that paper, is 64, thus it is significant to gain new independent Hubble parameters.

 This paper is organized as follows. We briefly give our LRGs selection sample in Section \ref{sec:sec}. In Section \ref{sec:third}, we describe explicitly how we gain the age information from the LRG spectra using \texttt{ULySS} and in Section \ref{sec:Four} we present our method to get the OHD from the galaxy ages.
 A cosmology constraint using all available OHD including our 4 new ones is given in Section \ref{sec:five}. Finally, in the last section, we discuss the limitation and prospect of our results.
\section{Source Selection of LRG Sample} \label{sec:sec}
It is important and necessary to select a large homogeneous passively evolving sample of LRGs to obtain the age-redshift
relation. The SDSS \citep[SDSS]{2000AJ....120.1579Y,2002AJ....123..485S,2003AJ....126.2081A} is currently the largest photometric and spectroscopic sky survey, which include five-band imagine over 10$^{4}$ deg$^{2}$ with accurate photometric calibration and spectroscopy of 10$^{6}$ galaxies\citep{2009ApJS..182..543A}. The SDSS spectroscopic survey consists of two samples of galaxies selected with different criteria, which are named the MAIN sample \citep{2002AJ....124.1810S} and the LRG sample \citep{2003ApJ...585..694E} respectively. The wavelength of these galaxy spectra cover the range from 3800\AA to 9200\AA with spectral resolution $\lambda/(\Delta\lambda)$ = 1850-2200, and with dedicated software the spectra are automatically reduced, which flux calibrates the spectra and references them to the heliocentric frame and converts to vacuum wavelengths.

\begin{figure}
\centering
\includegraphics[width=\textwidth, angle=0]{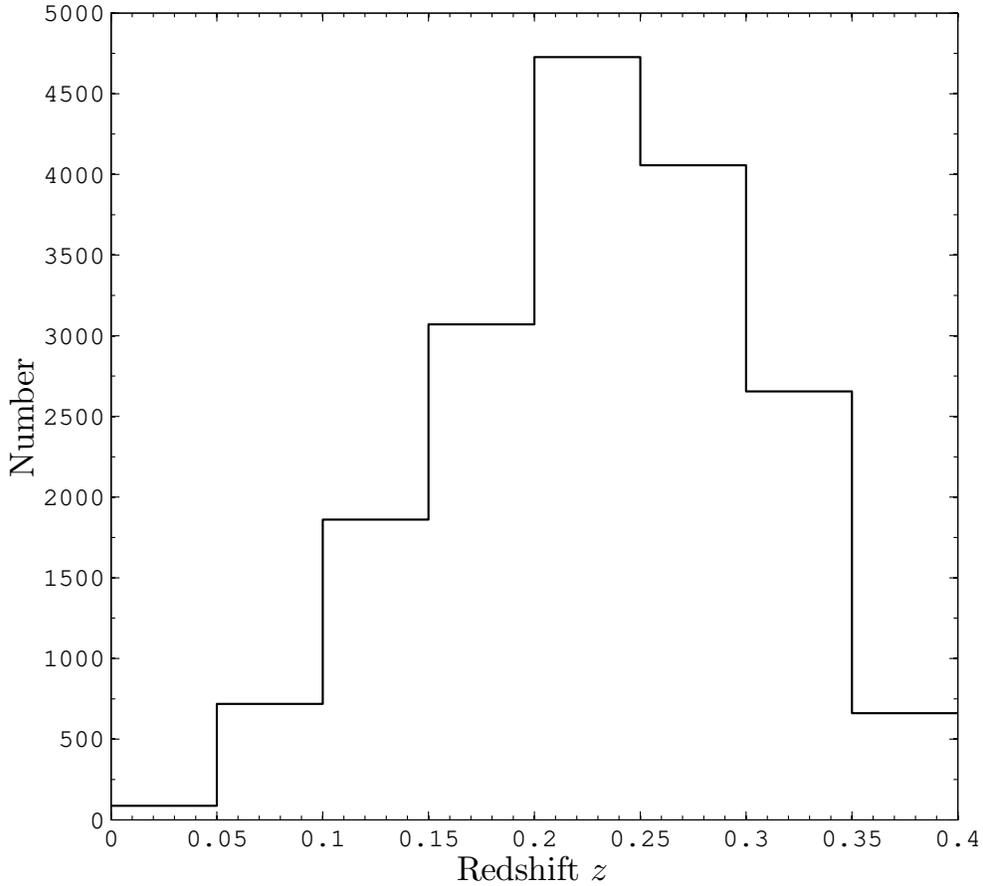}
\caption{Redshift distribution of 17832 LRGs.}\label{figure1}
\end{figure}

We chose the sample from LRG selection criterion of SDSS DR7, which is:
i)Selecting galaxies from the Catalog Archive Server (CAS) database with the \texttt{TARGET\_GALAXY\_RED} flag.
ii)The S/N of the $r$-waveband should be greater than 10.
iii)The restrictions, which are that \texttt{specClass = `SPEC\_GALAXY'}, that \texttt{zStar = `XCORR\_HIC'}, that \texttt{zWarning = 0}, that \texttt{eClass $<$ 0}, that \texttt{z $<$ 0.4} and that \texttt{fracDev\_r $>$ 0.8}, should also be satisfied.
The LRG selection criterion of SDSS \citep{2001AJ....122.2267E} is on the basis of color and magnitude to yield a sample of luminous, intrinsically red galaxies. However, the sample selected from SDSS according to Eisenstein's selection criterion are not very homogeneous.  Furthermore, we use the sample from Carson \& Nichol's sample \citep{2010MNRAS.408..213C} with a restraint on signal to noise ratio (SNR). As SNR has an impact on our next fitting step, it is necessary to demand SNR of  R-band to be greater than 10. The 17832 selected quiescent, luminous red galaxies basing on `Carson \& Nichol sample' method from SDSS DR7 cover the redshift range from 0-0.4. Redshift distribution of our sample is shown in Fig. \ref{figure1}.

Compared with other spectra selection work, `Carson \& Nichol sample' calibrates, for the first time, the SDSS spectra on Lick/Image Dissector Scanner (IDS) system. Their general selected criterion steps are as follows. First step is to get LRGs from Catalog Archive Server (CAS) as outlined in \citet{2001AJ....122.2267E}. Further, restraining the spectrum using standard emission lines such as H$\alpha$, H$\beta$ and OIII 5007 to increase the number of truly quiescent galaxies in the previous sample. To fully picture physical properties such as velocity dispersion and absolute luminosity of these LRGs, through correcting velocity dispersions for aperture effects and performing $K+e$ corrections to the magnitude, four subsamples
are produced with different absolute magnitude and velocity dispersion. For a explicit description of this method, please refer to \citet{2010MNRAS.408..213C}.

\section{Age-Redshift Relation } \label{sec:third}
We proceed to describe our way of obtaining age information of LRGs. There are many methods to learn the age and metallicity of stellar systems from a spectrum, such as SED fitting,
spectrophotometric indices (e.g., Lick, Rose indices) and full spectrum fitting \citep{2008MNRAS.385.1998K}. In this paper, we adopt the full spectrum fitting method to analyse the physical properties of stellar populations. Full spectrum fitting, which makes use of all the information contained in the signal, is insensitive to extinction or flux calibrating errors and independently from the shape of continuum. We adopt an open-source software, \texttt{ULySS}, to explore the history of stellar populations.

 \texttt{ULySS} is a full spectrum fitting package developed by a group at Universit$\acute{e}$ de Lyon \citep{2009A&A...501.1269K}. Its principle is to seek the minimum $\chi^2$ in the process of fitting an observed spectrum with a model spectrum in the pixel space with MPFIT function.  When fitting the observed spectrum ($F_{obs}(\lambda)$), the package uses a linear combination of $k$ non-linear components ($CMP_i$) with weight $W_i$ respectively to approximate it. In this process, the composite model is possibly convolved with a line-of-sight velocity distribution
 (LOSVD), multiplied by an $n^{th}$ order polynomial of $P_n(\lambda)$ and summed to another polynomial of $Q_m(\lambda)$ \citep[for more details please refer to][]{2009A&A...501.1269K}:
\begin{equation}
\label{eq:ulyss}
F_{obs}(\lambda)=P_n(\lambda)\times\{LOSVD(v_{sys},\sigma,h3,h4)\otimes\sum_{i=0}^{i=k}W_iCMP_i(a_1,a_2,a_3,...,\lambda)\}+Q_m(\lambda).
\end{equation}

For the study of stellar population, the CMP$_{i}$ is characterized by age and [Fe/H]. It uses Levenberg-Marquardt routine to evaluate parameters in individual CMP$_{i}$ and the coefficients of $P_n(\lambda)$ and $Q_m(\lambda)$ \citep{2009A&A...501.1269K}.
This method has already been successfully tested in \citet{2011A&A...525A..71W}. The reliability and robustness of using \texttt{ULySS} to the study of the history of stellar population has been verified
\citep[e.g.,][]{2009AN....330..960K,2009MNRAS.396.2133K}.

\subsection{Model Selection and Matching resolution}
There exist several population models and in our paper, three of them provided by the package \texttt{ULySS}, which are
Pegase-HR/ELODIE3.1, Galaxev/STELIB (hereafter BC03) and Vazdekis/Miles, are tested. The information of these models is listed in Tab. \ref{tab:mod}. \citet{2008MNRAS.385.1998K}also test these three models and verifies their reliability.

The full spectrum fitting uses the redundancy of the spectrum and the multiplicative polynomial could decrease the influence of flux calibration and Galactic extinction. On the other aspect, the fitting method is more sensitive to the wavelength range of the spectrum. The spectra from SDSS cover the wavelength from 3800\AA to 9200\AA, as we could see, only the wavelength range of BC03 could cover the whole wavelength of spectra from SDSS. After comparison of the three different models, we choose
BC03 as the reliable model for use.

\begin{table}[!h]
\tabcolsep 0pt
\vspace*{6pt}
\begin{center}
 \caption{Information in Three Models}\label{tab:mod}
\def\temptablewidth{1\textwidth}
{\rule{\temptablewidth}{1pt}}
\begin{tabular*}{\temptablewidth}{@{\extracolsep{\fill}}ccccccc}

  Model  &   Library  & Resolution /\AA& Wavelength /\AA&    Age/Gyr&   [Fe/H]/dex &        IMF \\
 \hline
    Pegase &  ELODIE3.1 &       0.55 &  4000-6800 &     0.1-20 &      -3.21-1.62 &  Salpeter\\

   Galaxev &     STELIB &          3 &  3200-9500 &     0.1-20 &      -2.3-0.4      &  Chabrier \\

  Vazdekis &      MILES &        2.3 &  3525-7500 &   0.1-17.5 &       -1.7-0.2 &   Salpeter \\
       \end{tabular*}
       {\rule{\temptablewidth}{1pt}}
       \end{center}
\vspace*{-8pt}

       \end{table}

 The first step in fitting is to match resolutions between the observed spectra with the model we chose. There are two ways for matching, either by transforming the resolution of the model or the observed spectrum. In our paper, we choose to transform the model provided by \texttt{ULySS} by injecting relative line spread function (LSF) between our spectrum and the model. When determining the LSF, we make use of the available velocity dispersion template stars as a standard star from http://www.sdss.org/dr7/algorithms/veldisp.html. We make a five times linear interpolation in wavelength and convolve with the model. In this way, a new matching model is generated, and all this process could be accomplished through \texttt{ULySS} function \citep{2009A&A...501.1269K}.

\subsection{Single Stellar Population Fit}
It is essential to study the stellar population of galaxies if we want to reconstruct the star formation history (SFH) of galaxies. We fit the spectrum with a single SSP for the following reasons. Firstly, LRG is believed to be drawn from a same parent population with the most percent of their stars are formed from a single-burst. Secondly, SSP-equivalent properties correspond to the `luminosity-weighted' average over the distributions. Thus by a single SSP fitting, a general view of a galaxy could be gotten \citep{2010MNRAS.409..567D}. We carry out an initial study of age-dating using the stellar population models of \citep{2003MNRAS.344.1000B,2010JCAP...02..008S} to synthesize spectra, which is also provided by \texttt{ULySS}.

 But when interpreting the galaxy spectra, we find both of age and metallicity guess value tend to affect the fitting result. It is difficult to remove such influence especially at the low resolving spectra \citep{2010MNRAS.409..567D}. The fitting of \texttt{ULySS} starts from a point (initial value) in the parameter space(age and metallicity), therefore we stress the importance of that point as the returning fitting value will be initial value dependent and the minimization will be local. In order to identify and understand the age-metallicity degeneracy and local minimizations, we analyse that age-metallicity degeneracy and construct $\chi^2$ map which is a function provided by \texttt{ULySS}. The $\chi^2$ map is a visualization of parameters space, and the principle of this map is to give a grid of nodes in a 2-D projection of the parameter space. $\chi^2$ map returns the minimization of each node \citep{2009A&A...501.1269K}. A subsample of 164 LRG spectra from redshift 0.03 to 0.179 is dealt with by building their $\chi^2$ maps. It has been discovered that the metallicity of LRGs are consistent with the assumption that selected galaxies are thought to be have similar metallicities \citep{2003ApJ...593..622J} and it floats between about 0.1 dex and 0.2 dex.

\begin{figure}
\includegraphics[width=\textwidth, angle=0]{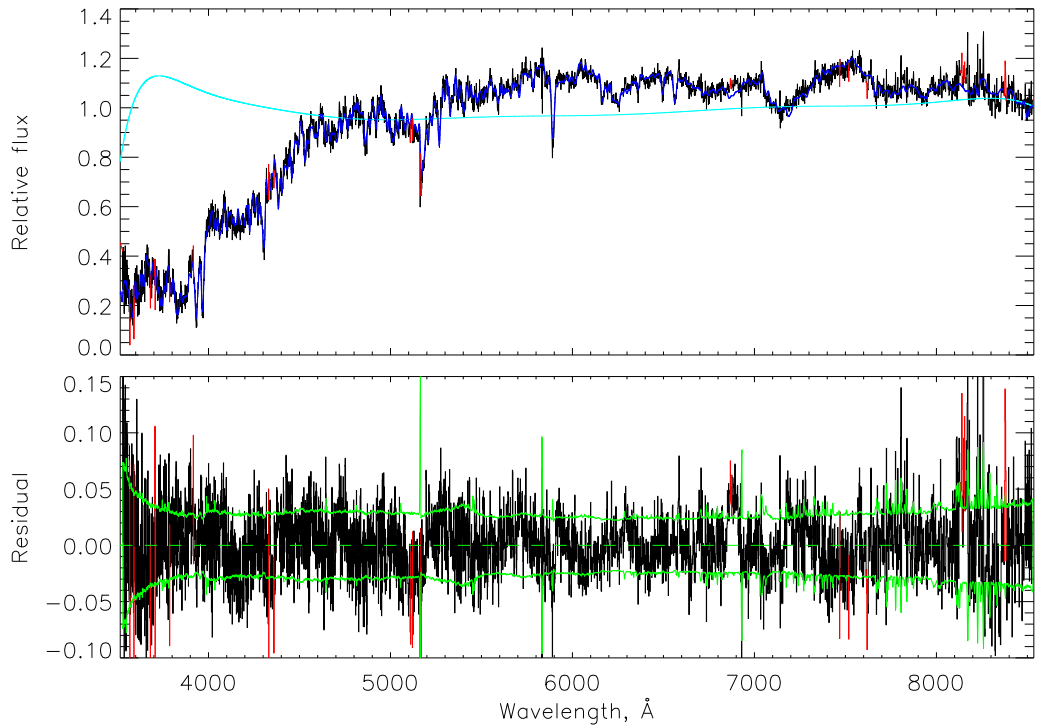}
\caption{Best fit with BC03 model and the residual spectrum for a galaxy. The top panel shows the spectrum in black and the best fit in blue. The red regions are rejected from the fit, the indigo line is the multiplicative polynomial. The bottom panel shows the residuals from the best fit. The continuous green lines represent the  1$\sigma$ deviation.}\label{fig:example}
\end{figure}

\texttt{ULySS} could provide initial value vectors as initial value, that is to say, if giving points intensive enough, it is possible to find the global minimization and thus break the degeneracy. We set our metallicity range guess points from 0.1-0.2 dex to every galaxy in our sample, while age guess point span from 5000 Myr-11200 Myr. Fig. \ref{fig:example} shows the best fit with BC03 model and the residual spectrum for a galaxy when a specific group of initial values is given.

\begin{figure}
\centering
\includegraphics[width=\textwidth]{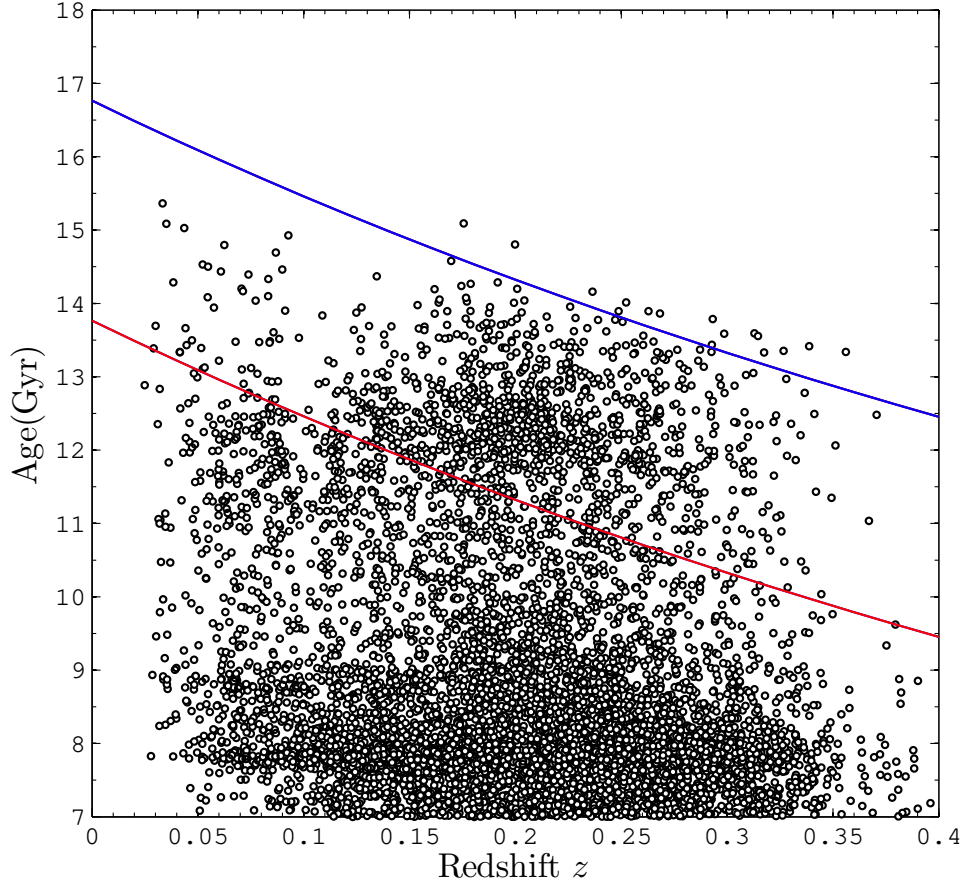}
\caption{Fitting results of the 17832 LRGs. For clarity, the ages less than 7Gyr have not been plotted. The red line shows the theoretical age of the universe $t(z)$ for a $\Lambda$CDM model with $\Omega_m=0.29$ and $H_0=69$ km\ s$^{-1}$ Mpc$^{-1}$ . The blue line indicates
$t(z)+3$Gyr, whose derivation of the 3Gyr comes from the systematic errors. A clear trend is present: the oldest ages of the galaxies decrease with redshift.}\label{fig:fz}
\end{figure}

  For each galaxy, groups of age and metallicity values can be gained as results. We ignore the metallicity and just focus on the age. To identify the final accurate age of each galaxy, we base our selections on the minimal $\chi^2$ criterion. In particular, if all fitting ages for a single galaxy surpass the age of our universe, it is not convincing as there are several reasons causing this situation such as the low resolution of spectra and model dependence.
   We set a limit of that $a+\sigma<16Gyr$, where the $a$ is the fitting age and  the $\sigma$ is the error, in the fitting process to exclude the situation that the age of a galaxy excess the age of the Universe. The fitting result is shown in Fig. \ref{fig:fz}. For clarity, the ages less than 7Gyr have not been plotted. In this figure, a clear tend that galaxies age decrease with the redshift is displayed.

\section{The determination of OHD} \label{sec:Four}
According to the Eq. \ref{eq:OHDe}, the slope of the linear fit of data of $t(z_i)$, which is the oldest age at redshift $z_i$, relates to the Hubble parameter directly. That is, the Hubble parameter at $z_{eff}$, $H(z_{eff})$, can be calculated by formula $H(z_{eff})=-[1/(1+z_{eff}](\Delta t/\Delta z)^{-1}$, where $z_{eff}=(z_{max}+z_{min})/2$.

Mathematically, the value of $\Delta z=z_{max}-z_{min}$ should not be too large, so we divide our data into four redshift regions,  $0.03\leq z \leq0.11$, $\ 0.08\leq z \leq0.16$, $0.16\leq z \leq0.24$, and $0.24\leq z \leq0.32$.

To calculate the $t(z_i)$ of each subsample, we adopt a common bin-dividing method. Take the first subsample for example. We divide it into several bins from 3 to 20, select the oldest galaxy of each bin as a set of $t(z_i)$ and fit the corresponding points of $t(z_i)$ with a straight line to get a candidate of OHD of this subsample. Here we get $18$ candidates of OHD of this subsample. The number of bins corresponding to the selected candidate is named as $n_{best}$. Three parameters are used here for selecting the most suitable number of bins that we choose: SSE$/n$ for the average of the Sum of Squares for Error,  $\sigma_{slope}/slope$ for the relative error of the fitting result of the slope and $P_{\geq\chi ^2}$ for the goodness of fit. The left subsamples follow the same method.

Here we explain our reason of choosing the above three parameters as criterion. Imagine that an envelop line can be obtained naturally with ideal data, therefore, we should take the bin-dividing number $n$ as large as possible. Unfortunately, for the real case, the envelop fluctuates strongly. This is due to the existence of `fake-oldest galaxies', which formed too late to be considered as `cosmic chronometer' compared with the LRGs at other redshifts. Therefore, large $n$ would increase the risk of selecting the `fake-oldest galaxies' as `cosmic chronometer'. That is what the parameter of SSE$/n$ can indicate, with value $0$ representing the ideal case. Conversely, too small $n$ represents large statistic errors, which is evaluated by parameter $\sigma_{slope}/slope$. Moreover, $P_{\geq\chi ^2}$ is calculated to represent the goodness of fit. Taking the three parameters into account, we finally obtain the value of the $n_{best}$ of each subsample.

In every subsample, we expect the $n_{best}$ satisfies the the smallest SSE$/n$, the largest $P_{\geq\chi ^2}$ and the smallest absolute value of $\sigma_{slope}/slope$.
In the first subsample, $n=6$ case meet the criterion perfectly. So does $n=7$ in the second subsample and $n=6$ in the third subsample.  For the fourth case, judging the $\sigma_{slope}/slope$  first, we can see for both $n=10$ and $n=12$ share the least value. Then, considering the second parameter SSE$/n$, we chose $n=12$ for  smaller value. Besides, its $P_{\geq\chi ^2}$ is also satisfied.

\begin{figure}
\includegraphics[width=\textwidth, angle=0]{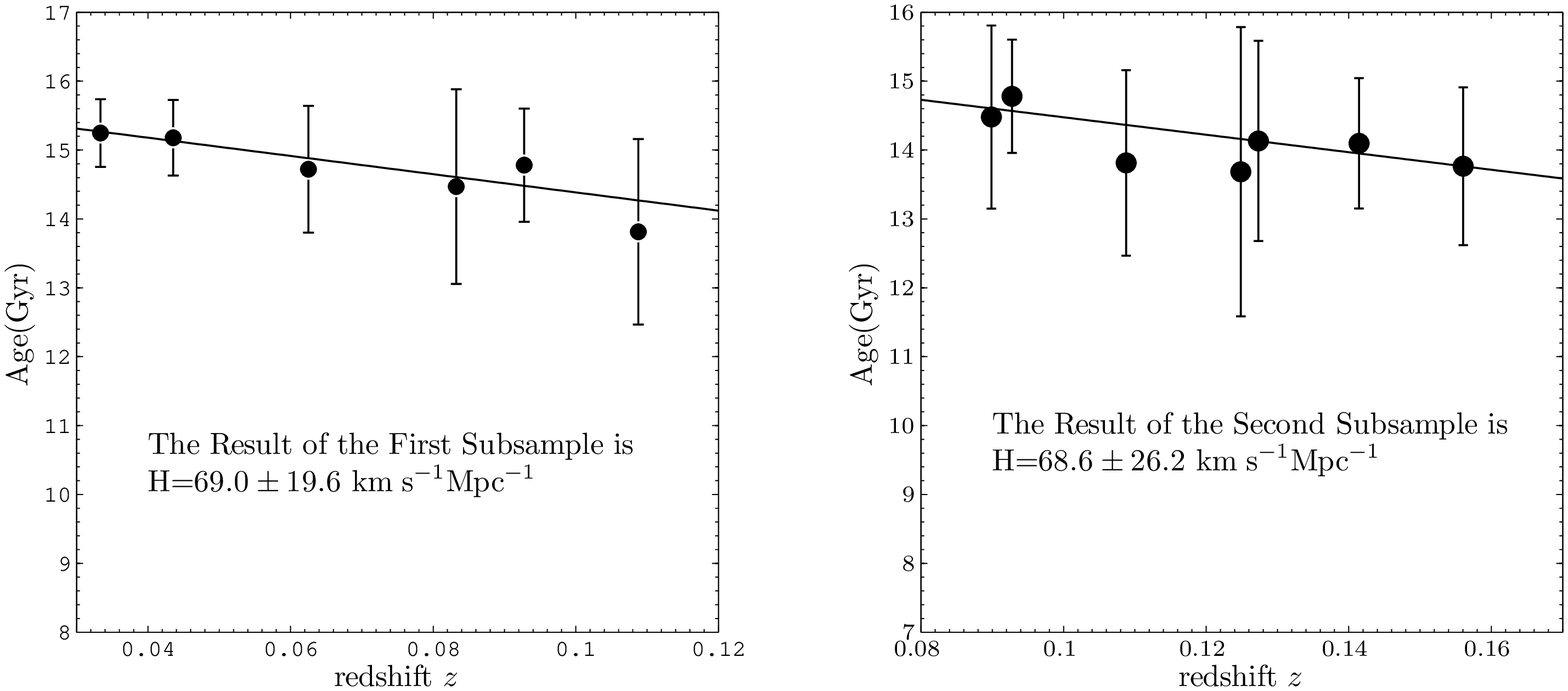}
\includegraphics[width=\textwidth, angle=0]{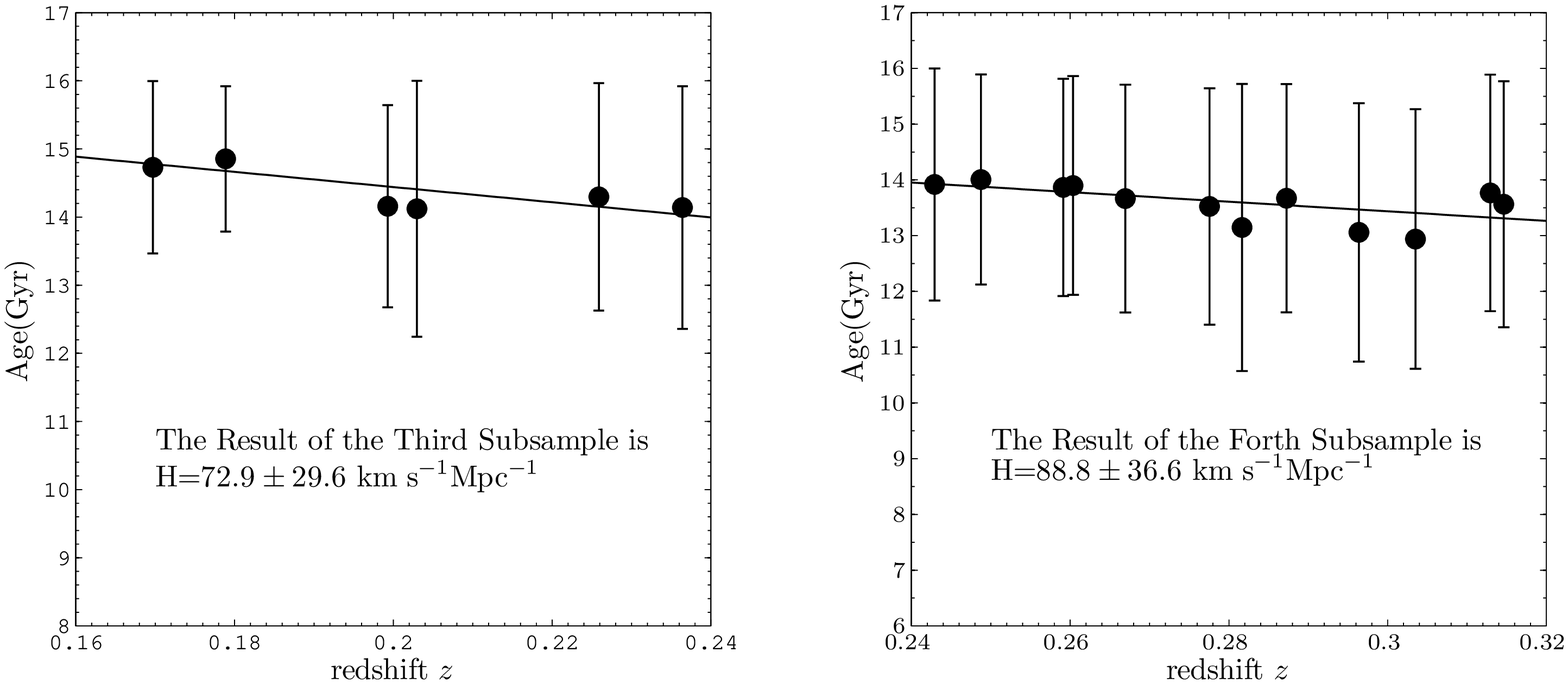}
\caption{The oldest ages in each bin when we divide the subsamples into their corresponding $n_{best}$ bins and their optimal fitted results. The solid line represents the best fitting for each subsample.  In the first subsample $z_{min}=0.033$, $z_{max}=0.109$, and $z_{eff}=0.07$. In the second subsample, $z_{min}=0.090$, $z_{max}=0.156$, $z_{eff}=0.12$. In the third subsample, $z_{min}=0.170$, $z_{max}=0.236$, $z_{eff}=0.20$, and in the forth subsample, $z_{min}=0.243$, $z_{max}=0.315$, $z_{eff}=0.28$.}\label{fig:fitres}
\end{figure}

Fig.\ref{fig:fitres} shows the oldest ages in each bin when we divide the subsamples into their corresponding $n_{best}$ bins and their optimal fits are also plotted.

Then we will introduce how we get the best parameters of linear fitting and their error bars.

When we fit $n$ set of data ($z_i$, $t_i$) with a straight line $t=kz+b$ and the error of $t_i$ is $\sigma_i$ ($i=1,2,...,n$), the best parameters of the linear fitting are obtained by minimizing the $\chi^2$:
 \begin{equation}\label{eq:chioflinearfit}
   \chi^2=\sum_{i=1}^{n}\frac{(t_i-kz_i-b)^2}{\sigma_i^2},
 \end{equation}
To get the error of $\sigma_k$, we must rewrite the Eq. \ref{eq:chioflinearfit} as follows so that we can use some formulae of the regressive parameter errors:
 \begin{equation}
   \chi^2=\sum_{i=1}^{n}({\frac{t_i}{\sigma_i}-k\frac{z_i}{\sigma_i}-b\frac{1}{\sigma_i}})^2.
 \end{equation}
 This form of $\chi^2$ is same as that when we fit the data of ($t_i/\sigma_i$,$z_i/\sigma_i, 1/\sigma_i$) with linear function of $\frac{t}{\sigma}=k\frac{z}{\sigma}+b\frac{1}{\sigma}$ with the same method of minimum $\chi^2$.

Then, using the well known formula on confidence intervals of regression coefficients at a confidence level of $1-\alpha$ \citep[seeing any textbook on regression such as][for detail]{regression}, we can get the error of the $k$ which can be regarded as $\sigma_{k}$ (i.e. the $\sigma_{slope}$ in Tab. \ref{tab:fitres}):
 \begin{equation}\label{eq:getsigmak}
  \sigma_k= \sigma_{slope}=|t_{(\alpha/2,\ n-2)}|\sqrt{\frac{\textrm{S(1,1)}}{n-2}\times \textrm{SSE}},
 \end{equation}
 where $|t_{(\alpha/2,\ n-2)}|$ is the absolute value of the inverse of Student's t-CDF (Cumulative Distribution Function) with degrees of freedom $n-2$ for the corresponding probabilities in $\alpha/2$, which can be calculated with function of tinv in MATLAB, SSE is the Sum of Squares for Error and the $S(1,1)$ is the element in the first row and the first column in the inverse of the following matrix:
 \begin{equation}
 \left(
\begin{array}{cc}
 \sum\limits_{i=1}^{n}\frac{z_i^2}{\sigma_i^2},\ \  & \sum\limits_{i=1}^{n}\frac{z_i}{\sigma_i^2} \\
 \\
 \sum\limits_{i=1}^{n}\frac{z_i}{\sigma_i^2},\ \  & \sum\limits_{i=1}^{n}\frac{1}{\sigma_i^2}
\end{array}
\right).
\end{equation}
In addition, the steps above can be completed by MATLAB Toolboxes easily.

With the Eq. \ref{eq:OHDe}, we can get the relation between the error of $H(z)$ and the $\sigma_{slope}$:
\begin{equation}
\sigma_H=\frac{1}{1+z_{eff}}\frac{1}{\sigma_{slope}^2},
\end{equation}
with which the error of $H(z)$ is calculated finally.

\begin{table}[!h]
\tabcolsep 0pt
\vspace*{6pt}
\begin{center}
 \caption{Fitting results}\label{tab:fitres}
\def\temptablewidth{1\textwidth}
{\rule{\temptablewidth}{1pt}}
\begin{tabular*}{\temptablewidth}{@{\extracolsep{\fill}}c|c c c c c | c c c c c}
$n$ &4&5&6&7&8&5&6&7&8&9\\ \hline
 $-\sigma_{slope}/slope$ & 0.64 & 0.44 & 0.28 & 0.31 & 0.28 &0.49&1.45 &0.38 &0.54 &0.74\\
SSE$/n$ & 0.053 & 0.076 & 0.049 & 0.106 & 0.094 & 0.251 & 0.964 & 0.319 & 0.802 & 1.575\\
$P_{\geq\chi ^2}$\dag& 0.900 &	0.945 & 0.990 & 0.980 & 0.993& 0.969 & 0.915 & 0.997 & 0.992 & 0.980\\
  \hline

  $n$ & 4 & 5 & 6 & 7 & 8 &10 & 11 & 12 & 13 & 14\\ \hline
  $-\sigma_{slope}/slope$ & 1.08 & 0.43 & 0.41 & 0.50 & 0.44 & 0.41 & 0.43 & 0.41 & 0.43 & 0.43\\
   SSE$/n$ & 0.028 & 0.019 & 0.017 & 0.034 & 0.031&0.208&0.113 & 0.175 & 0.163 & 0.211 \\
    $P_{\geq\chi ^2}$\dag& 0.945 & 0.992 & 0.999 & 0.999 & 0.999&  1.000 & 1.000 & 1.000 & 1.000 & 1.000 \\
     \hline
       \end{tabular*}
       {\rule{\temptablewidth}{1pt}}
       \end{center}
\vspace*{-8pt}
 \small \dag Too big $P_{\geq\chi ^2}$ in these two subsamples result from too great errors. The two other parameters \newline are taken account of mainly because of the very close $P_{\geq\chi ^2}$ in this subsample.
       \end{table}

\begin{figure}
\centering
\includegraphics[width=\textwidth, angle=0]{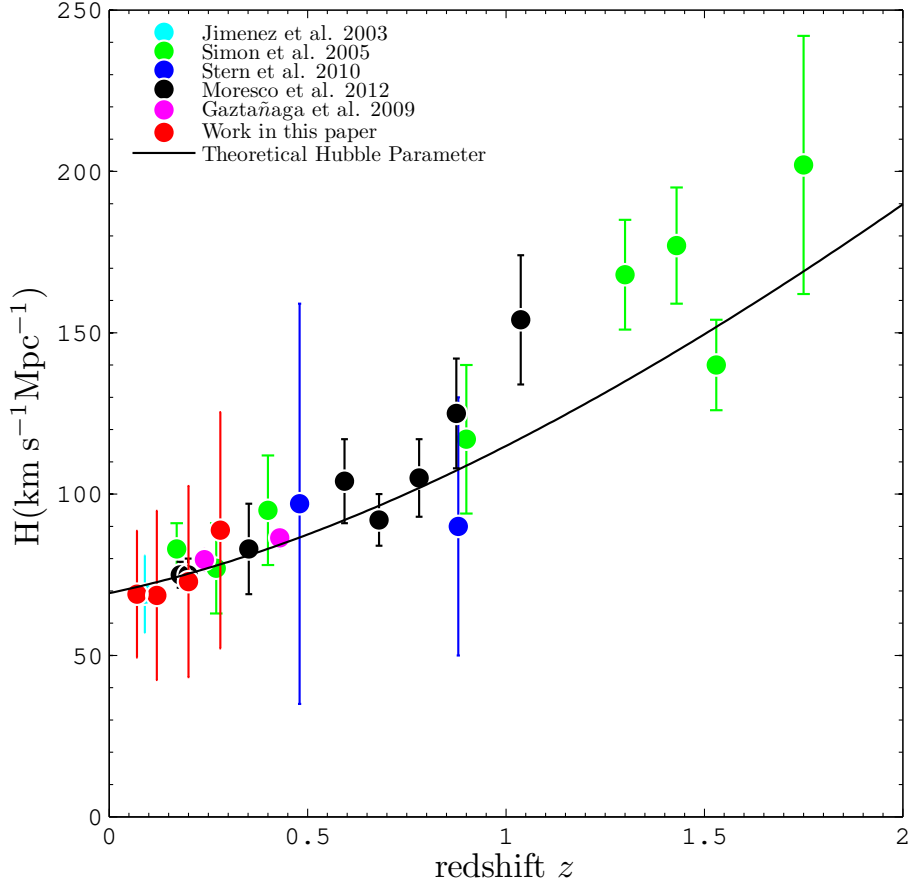}
\caption{All available OHD points. The solid  line plots the theoretical Hubble parameter $H_{fid}$ as a function of $z$ from the spatially flat
$\Lambda$CDM model with $\Omega_m = 0.3$, $\Omega_{\Lambda} = 0.7$, and $H_0$ = 72\ km s$^{-1}$Mpc$^{-1}$. The OHD points are listed in Tab. \ref{tab:AOHD}}\label{figure6}
\end{figure}

\begin{table}[!h]
\tabcolsep 0pt
\vspace*{-10pt}
\begin{center}
 \caption{All available OHD}\label{tab:AOHD}
\def\temptablewidth{1.0\textwidth}
{\rule{\temptablewidth}{1pt}}
\begin{tabular*}{\temptablewidth}{@{\extracolsep{\fill}}cccc}
 $z$   &$H(z)^\dag$     &$\sigma_{H(z)^\dag}$&Ref.\\\hline
    0.090  & 69    & 12    &\cite{2003ApJ...593..622J} \\
    0.170 & 83    & 8     & \cite{2005PhRvD..71l3001S} \\
    0.270 & 77    & 14    & \cite{2005PhRvD..71l3001S} \\
    0.400  & 95    & 17    & \cite{2005PhRvD..71l3001S} \\
    0.900 & 117   & 23    & \cite{2005PhRvD..71l3001S} \\
    1.300 & 168   & 17    & \cite{2005PhRvD..71l3001S} \\
    1.430 & 177   & 18    & \cite{2005PhRvD..71l3001S} \\
    1.530 & 140   & 14    & \cite{2005PhRvD..71l3001S} \\
    1.750 & 202   & 40    & \cite{2005PhRvD..71l3001S} \\
    0.480 & 97    & 62    & \cite{2010JCAP...02..008S} \\
    0.880 & 90    & 40    & \cite{2010JCAP...02..008S} \\
    0.179 & 75    & 4     & \cite{2012JCAP...08..006M} \\
    0.199 & 75    & 5     & \cite{2012JCAP...08..006M} \\
    0.352 & 83    & 14    & \cite{2012JCAP...08..006M} \\
    0.593 & 104   & 13    & \cite{2012JCAP...08..006M} \\
    0.680 & 92    & 8     & \cite{2012JCAP...08..006M}\\
    0.781 & 105   & 12    & \cite{2012JCAP...08..006M} \\
    0.875 & 125   & 17    & \cite{2012JCAP...08..006M} \\
    1.037 & 154   & 20    & \cite{2012JCAP...08..006M} \\
    0.24  &79.69  &3.32   & \cite{2009MNRAS.399.1663G}\\
    0.43  &86.45  &3.27   & \cite{2009MNRAS.399.1663G} \\
    0.07  & 69.0  & 19.6  &   \dag\dag \\
    0.12  & 68.6  & 26.2  &   \dag\dag     \\
    0.20  & 72.9  & 29.6  &   \dag\dag   \\
    0.28  & 88.8  & 36.6  &   \dag\dag      \\ \hline
       \end{tabular*}
       {\rule{\temptablewidth}{1pt}}
       \end{center}
       \vspace*{-8pt}
        \small \dag\ \   The unit is km s$^{-1}$Mpc$^{-1}$. \\ \dag\dag\  Work in this paper.\\
       \end{table}

\section{Cosmological constraints from OHD} \label{sec:five}

Now we have totally 25 available OHD which are listed in Tab. \ref{tab:AOHD}. Using these 25 OHD, we constrain the cosmological parameters. The best fit parameters of the model via OHD are determined by minimizing the $\chi^2$:
\begin{equation}
\label{eq:chi2}
\chi ^2_{OHD} (\mathbf{p_{model}}) = \sum\limits_{i = 1}^{N} {\frac{(H_{obs} (z_i ) -
H_{th} (z_i ))^2}{\sigma _{obs,i} ^2}},
\end{equation}
where $\mathbf{p_{model}}$ is a free parameter.
Based on the basic equations of $\Lambda$CDM model, we have the following two models: for a flat $\Lambda$CDM($\Omega_k=0$),
 $H_{th}(z)  = H_0\sqrt{\Omega _m(1+z)^3+(1-\Omega _m)}$ with $\mathbf{p_{flat}} = (\Omega _m ,H_0 )$; for a non-flat $\Lambda$CDM,
$ H_{th}(z)  = H_0\sqrt{\Omega _m(1+z)^3+\Omega_\Lambda+(1-\Omega _\Lambda-\Omega_m)(1+z)^2}$,
with $\mathbf{p_{non-flat}} = (\Omega_\Lambda,\Omega _m ,H_0 )$.
For both models, the likelihood function can be written as $L \propto \exp$ $(-\chi_{OHD}^2 / 2)$.

We use MCMC method to calculate the likelihood in the independent parameter space. The Markov chains are generated and analysed via the Python MCMC code -- \texttt{Pymc}. The parameter $H_0$ and the density parameter $\mathbf{\Omega}=(\Omega_m,\  \Omega_\Lambda)$ are treated as independent parameters in these two models \citep{2010PhLB..687..286W}, and all of the prior distributions of them are set as the uniform distribution (Tab. \ref{tab:tpoimp}).

\begin{table}[!h]
\tabcolsep 0pt
\vspace*{-10pt}
\begin{center}
\caption{The Prior of Model Parameters}\label{tab:tpoimp}
\def\temptablewidth{0.7\textwidth}
{\rule{\temptablewidth}{1pt}}
\begin{tabular*}{\temptablewidth}{@{\extracolsep{\fill}}cc}
 Model Parameters & Prior Distribution\\ \hline
  $\Omega_m$ &  Uniform(0.0,1.5)\dag\\
 $\Omega_{\Lambda}$ &  Uniform(0.0,2.5)\dag\\
 $H_0$  &   Uniform(50,100) \dag\\
       \end{tabular*}
       {\rule{\temptablewidth}{1pt}}
       \end{center}
       \vspace*{-8pt}
\small \dag\emph{Uniform(lower limit,upper limit)} stand for an uniform distribution in the interval [lower limit,upper limit], and the same below.
       \end{table}

\begin{figure}
\includegraphics[width=\textwidth, angle=0]{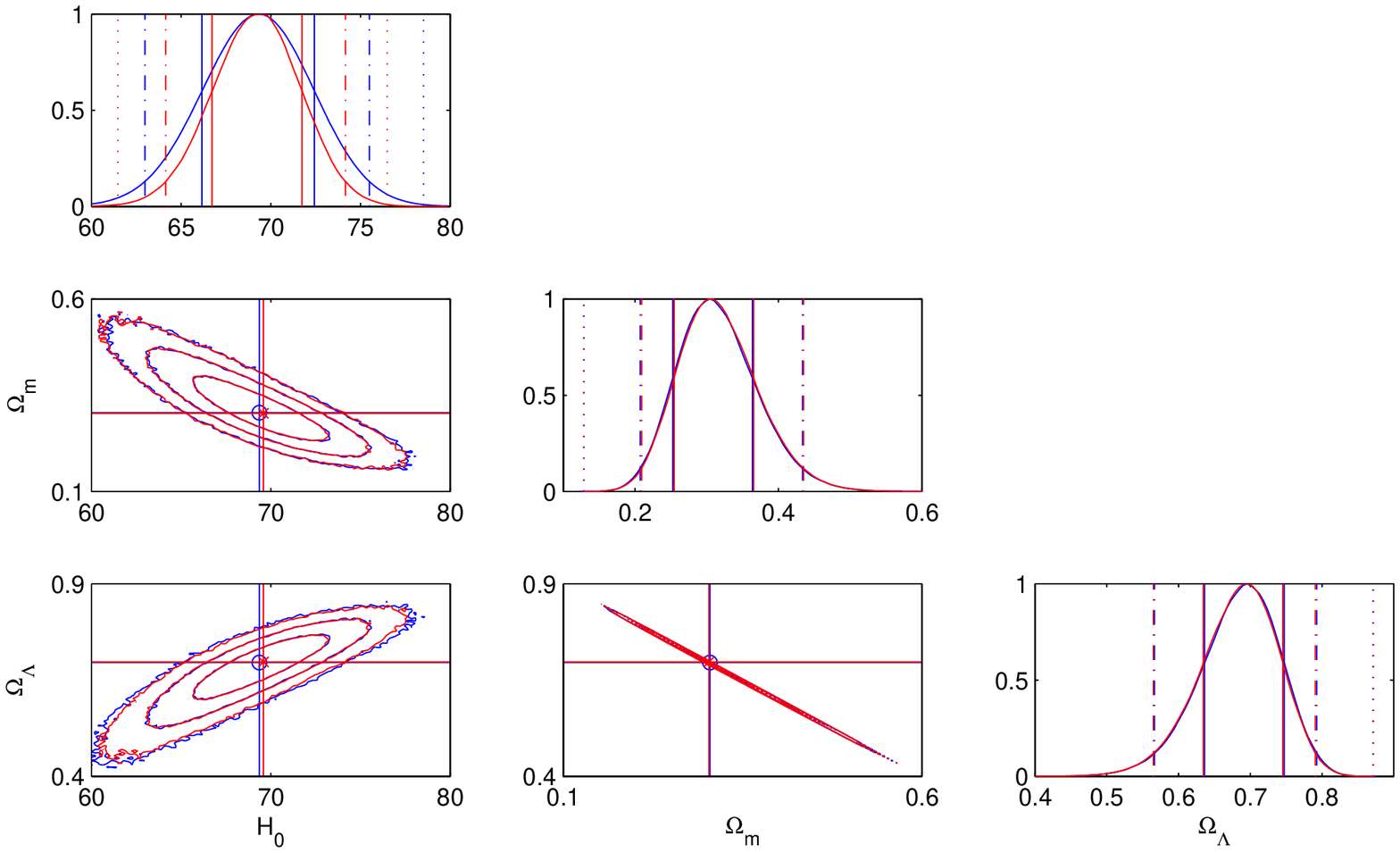}
  \caption{OHD constraint region of flat $\Lambda$CDM. Red lines represent the `new version' of OHD, with the blue ones for the `old version'. The diagonal plots show the 1-D PDF and the vertical lines from center to edge match the $3\sigma$-s intervals; The non-diagonal plots show the 2-D confidence regions with contours for inner to outer match the $3\sigma$-s levels, and the open blue circle points out the best fit point by `old version' while the red cross shows the ones via `new version'. 
}
  \label{fig:soccf}
\end{figure}

\begin{figure}
\includegraphics[width=\textwidth, angle=0]{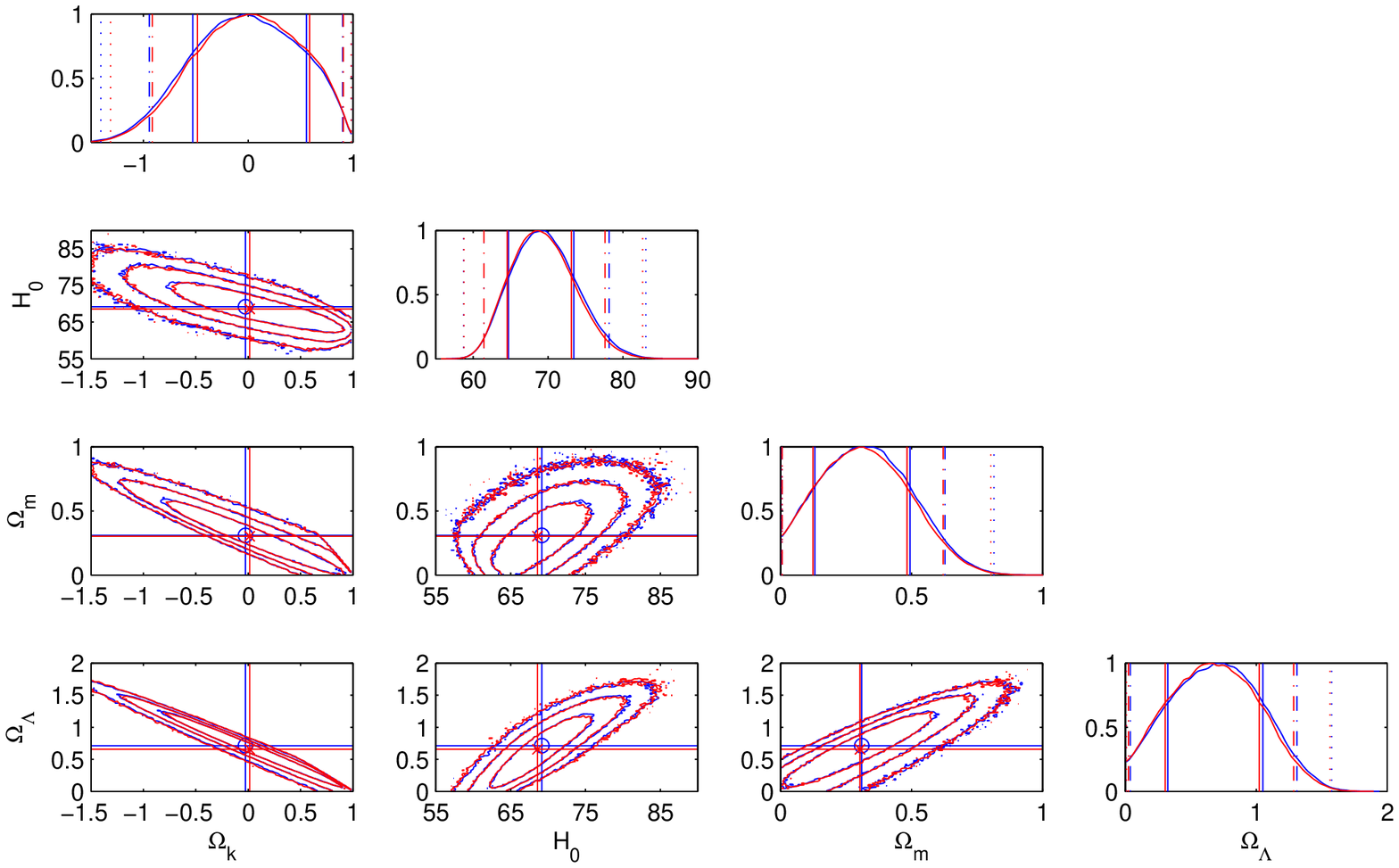}
  \caption{ OHD constraint region of non-flat $\Lambda$CDM. Red lines represents the `new version' of OHD, with blue ones for the `old version'. The diagonal plots show the 1-D PDF and the vertical lines from center to edge match the $3\sigma$-s intervals; The non-diagonal plots show the 2-D confidence regions with contours for inner to outer match the $3\sigma$-s levels, and the open blue circle points out the best fit point by `old version' while the red cross shows the ones via `new version'. 
 }
 \label{fig:soccs}
\end{figure}

Fig. \ref{fig:soccf} and Fig. \ref{fig:soccs} show the one-dimensional marginalized probability distribution for each parameter at the diagonal positions and the two-dimensional marginalized confidence regions at the non-diagonal positions of the flat $\Lambda$CDM model and non-flat $\Lambda$CDM model respectively. The $3 \sigma$-s confidence regions ($ 68.7{\%},\ 95.45{\%},\ 99.73{\%} $ ) for each parameter are calculated carefully. The best fit parameters are
pointed out in the contour plots with a long vertical line and a long horizontal line across.

To compare the constraining results of the `new version OHD' with the new points we obtained in this paper added and the `old version OHD' (see Tab. \ref{tab:AOHD} for detail), we display both results of the same model in Fig. \ref{fig:soccf} and Fig. \ref{fig:soccs} with different line colors to distinguish: the red lines referring to the `new version OHD' and the blue ones means the `old version OHD'.

The constraints in Fig. \ref{fig:soccf} and Fig. \ref{fig:soccs} show the well performances on constraining $\Lambda$CDM. In Fig \ref{fig:soccf} and Fig \ref{fig:soccs} the 1-D marginalized constraints are more stringent than the old one and for the same model the contour plots against the same parameters pair have more shrunk confidence region at the same level. Since the number of new added-on points are so small, the degree of shrinking is weak as well.

{\renewcommand\baselinestretch{2}\selectfont

\par}

\section{Conclusion and Discussion}
In this paper, we present our measurements of four new OHD data points from the ages of passively-evolving galaxies at redshift $0<$z$<0.4$.
A large sample of LRGs spectra has been fitted by us with SSP models and a age-redshift relation is obtained and displayed. By computing the relative ages of these LRGs, we gain four new OHD data points. Combining them with other available 21 OHD points, we constrain cosmological parameters using the updated dataset of OHD. It should be mentioned the similar work  \cite{2012ApJ...758..107L}, in which they used the SDSS DR7 to constrain the $H_0$, a particular Hubble parameter at $z=0$. We hope to give a tighter constraint on cosmology parameters with our new OHD data points. Unfortunately, these four points do limited improvements for constraining cosmological parameters more accurately, because of the relatively large error bar of these points.

Here, we explain the possible reasons. Firstly, the low SNR of spectra from SDSS leads to the large uncertainty when calculating the age of the galaxies. Combining spectra from various observation project may be a solution, as discussed in \cite{2003ApJ...593..622J,2010JCAP...02..008S}. Comparing our points with the previous, especially those in \cite{2005PhRvD..71l3001S} and \cite{2012JCAP...08..006M}, although there really exists a big difference in terms of the accuracy, we would like to illustrate as followings. One is a theoretically good method to obtain the age of LRG in \cite{2012JCAP...08..006M}. Still, there are many problems in the method to be solved. The other is the results in \cite{2005PhRvD..71l3001S} with more accurate than ours because of the high quality of their spectra. OHD data points in \cite{2010JCAP...02..008S}, whose quality of spectrum and method of fitting spectra are similar to ours, share the comparable degree of accuracy with us, which also proves the objectivity of our OHD data points.

Second, as our selected LRGs are incomplete and
unable to cover all old-age galaxies in the universe, there would be some difficulty in tracing the `cosmic chronometers'. That is to say, oldest age in our sample may not represent cosmic age at that redshift. It would be improved through future redshift surveys of observation and lead to successful realization of differential age method. In addition, as high SNR is essential for a precise fitting which determines the age of galaxies, we suggest that the accuracy of OHD would be improved if the SNR of spectra would increase. We employ \texttt{ULySS} to reconstruct the stellar population of galaxies. Though the robustness of \texttt{ULySS} has been illustrated, the fitting minimal may still be local because of the limitation of groups of initial values. There has been significant advancement in modelling stellar populations of LRG galaxies and we expect these to improve the accuracy in this process in the future.

The number of OHD is still scarce compared with SNIa data sets.
Advantages of constraining cosmological models with the OHD \citep{2002ApJ...573...37J,2001PhRvL..86....6M} are proved, therefore increasing the number of OHD is imperative. OHD plays almost the same role as that of SNIa for the joint constraints on the $\Lambda$CDM model. The number of OHD points would extended in further decades with more and deeper observation of galaxies and at that time OHD set lone is capable to be used in place of current SNIa data sets \citep{2011ApJ...730...74M}. Fortunately, we have seen that both project and proposed Sandage-Loeb observational plan \citep{2007PhRvD..75f2001C} can be used to extend our knowledge of cosmic expansion into the even deeper redshift. Finally, we think it is reasonable to expect that OHD will complement SNIa, BAO and weak lensing and help us detect more information of the evolution history of our universe.

ACKNOWLEDGMENTS. Special thanks to Robert C. Nichol and Dan P. Carson for providing us the LRG sample data. We are grateful to G. Bruzual, M. Koleva, Wei Du, Gao-Chao Liu, Xian-Min Meng, Xue-Lei Chen, You-Jun Lu and Cong Ma for their helpful suggestions. 
This work
was supported by the National Science Foundation of China
(Grants No. 11173006), the Ministry of Science and Technology National Basic Science program (project 973) under grant
No. 2012CB821804, and the Fundamental Research Funds for
the Central Universities.

\clearpage

\bibliography{ms1524}

\begin{thebibliography}{41}
\providecommand{\natexlab}[1]{#1}
\providecommand{\selectlanguage}[1]{\relax}

\bibitem[{{Abazajian} et~al.(2003)}]{2003AJ....126.2081A}
{Abazajian}, K., et~al. 2003, \aj, 126, 2081

\bibitem[{{Abazajian} et~al.(2009){Abazajian}, {Adelman-McCarthy},
  {Ag{\"u}eros} et~al.}]{2009ApJS..182..543A}
{Abazajian}, K.~N., {Adelman-McCarthy}, J.~K., {Ag{\"u}eros}, M.~A., et~al.
  2009, \apjs, 182, 543

\bibitem[{{Bruzual} \& {Charlot}(2003)}]{2003MNRAS.344.1000B}
{Bruzual}, G., \& {Charlot}, S. 2003, \mnras, 344, 1000

\bibitem[{{Carson} \& {Nichol}(2010)}]{2010MNRAS.408..213C}
{Carson}, D.~P., \& {Nichol}, R.~C. 2010, \mnras, 408, 213

\bibitem[{{Chen} \& {Ratra}(2011)}]{2011PhLB..703..406C}
{Chen}, Y., \& {Ratra}, B. 2011, Physics Letters B, 703, 406

\bibitem[{{Corasaniti} et~al.(2007){Corasaniti}, {Huterer}, \&
  {Melchiorri}}]{2007PhRvD..75f2001C}
{Corasaniti}, P.-S., {Huterer}, D., \& {Melchiorri}, A. 2007, \prd, 75, 062001

\bibitem[{{Crawford} et~al.(2010)}]{2010MNRAS.406.2569C}
{Crawford}, S.~M., et~al. 2010, \mnras, 406, 2569

\bibitem[{{Du} et~al.(2010){Du}, {Luo}, {Prugniel}, {Liang}, \&
  {Zhao}}]{2010MNRAS.409..567D}
{Du}, W., {Luo}, A.~L., {Prugniel}, P., {Liang}, Y.~C., \& {Zhao}, Y.~H. 2010,
  \mnras, 409, 567

\bibitem[{{Eisenstein} et~al.(2003){Eisenstein}, {Hogg}
  et~al.}]{2003ApJ...585..694E}
{Eisenstein}, D.~J., {Hogg}, D.~W., et~al. 2003, \apj, 585, 694

\bibitem[{{Eisenstein} et~al.(2005){Eisenstein}, {Zehavi}
  et~al.}]{2005ApJ...633..560E}
{Eisenstein}, D.~J., {Zehavi}, I., et~al. 2005, \apj, 633, 560

\bibitem[{{Eisenstein} et~al.(2001)}]{2001AJ....122.2267E}
{Eisenstein}, D.~J., et~al. 2001, \aj, 122, 2267

\bibitem[{{Gazta{\~n}aga} et~al.(2009){Gazta{\~n}aga}, {Cabr{\'e}}, \&
  {Hui}}]{2009MNRAS.399.1663G}
{Gazta{\~n}aga}, E., {Cabr{\'e}}, A., \& {Hui}, L. 2009, \mnras, 399, 1663

\bibitem[{{Ghirlanda} et~al.(2004){Ghirlanda}, {Ghisellini}, {Lazzati}, \&
  {Firmani}}]{2004ApJ...613L..13G}
{Ghirlanda}, G., {Ghisellini}, G., {Lazzati}, D., \& {Firmani}, C. 2004, \apjl,
  613, L13

\bibitem[{{He} \& {Liu}(2011)}]{regression}
{He}, X., \& {Liu}, W. 2011, {Applied regression analysis} (China Renmin
  University Press)

\bibitem[{{Hicken} et~al.(2009)}]{2009ApJ...700.1097H}
{Hicken}, M., et~al. 2009, \apj, 700, 1097

\bibitem[{{Jimenez} \& {Loeb}(2002)}]{2002ApJ...573...37J}
{Jimenez}, R., \& {Loeb}, A. 2002, \apj, 573, 37

\bibitem[{{Jimenez} et~al.(2003){Jimenez}, {Verde}, {Treu}, \&
  {Stern}}]{2003ApJ...593..622J}
{Jimenez}, R., {Verde}, L., {Treu}, T., \& {Stern}, D. 2003, ApJ, 593, 622

\bibitem[{{Koleva} et~al.(2009{\natexlab{a}}){Koleva}, {de Rijcke}, {Prugniel},
  {Zeilinger}, \& {Michielsen}}]{2009MNRAS.396.2133K}
{Koleva}, M., {de Rijcke}, S., {Prugniel}, P., {Zeilinger}, W.~W., \&
  {Michielsen}, D. 2009{\natexlab{a}}, \mnras, 396, 2133

\bibitem[{{Koleva} et~al.(2009{\natexlab{b}}){Koleva}, {Prugniel}, {Bouchard},
  \& {Wu}}]{2009A&A...501.1269K}
{Koleva}, M., {Prugniel}, P., {Bouchard}, A., \& {Wu}, Y. 2009{\natexlab{b}},
  \aap, 501, 1269

\bibitem[{{Koleva} et~al.(2009{\natexlab{c}}){Koleva}, {Prugniel}, {De Rijcke},
  {Zeilinger}, \& {Michielsen}}]{2009AN....330..960K}
{Koleva}, M., {Prugniel}, P., {De Rijcke}, S., {Zeilinger}, W.~W., \&
  {Michielsen}, D. 2009{\natexlab{c}}, Astronomische Nachrichten, 330, 960

\bibitem[{{Koleva} et~al.(2008){Koleva}, {Prugniel}, {Ocvirk}, {Le Borgne}, \&
  {Soubiran}}]{2008MNRAS.385.1998K}
{Koleva}, M., {Prugniel}, P., {Ocvirk}, P., {Le Borgne}, D., \& {Soubiran}, C.
  2008, \mnras, 385, 1998

\bibitem[{{Komatsu} et~al.(2011)}]{2011ApJS..192...18K}
{Komatsu}, E., et~al. 2011, \apjs, 192, 18

\bibitem[{{Li} et~al.(2008)}]{2008ApJ...680...92L}
{Li}, H., et~al. 2008, \apj, 680, 92

\bibitem[{{Li} et~al.(2009){Li}, {Li}, {Wang}, \&
  {Zhang}}]{2009JCAP...06..036L}
{Li}, M., {Li}, X.-D., {Wang}, S., \& {Zhang}, X. 2009, \jcap, 6, 036

\bibitem[{{Liu} et~al.(2012){Liu}, {Lu}, {Chen} et~al.}]{2012ApJ...758..107L}
{Liu}, G., {Lu}, Y., {Chen}, X., et~al. 2012, \apj, 758, 107

\bibitem[{{Ma} \& {Zhang}(2011)}]{2011ApJ...730...74M}
{Ma}, C., \& {Zhang}, T.-J. 2011, \apj, 730, 74

\bibitem[{{Maor} et~al.(2001){Maor}, {Brustein}, \&
  {Steinhardt}}]{2001PhRvL..86....6M}
{Maor}, I., {Brustein}, R., \& {Steinhardt}, P.~J. 2001, Physical Review
  Letters, 86, 6

\bibitem[{{Moresco} et~al.(2012)}]{2012JCAP...08..006M}
{Moresco}, M., et~al. 2012, \jcap, 8, 006

\bibitem[{{Percival} et~al.(2010)}]{2010MNRAS.401.2148P}
{Percival}, W.~J., et~al. 2010, \mnras, 401, 2148

\bibitem[{{Riess} et~al.(1998)}]{1998AJ....116.1009R}
{Riess}, A.~G., et~al. 1998, \aj, 116, 1009

\bibitem[{{Roseboom} et~al.(2006)}]{2006MNRAS.373..349R}
{Roseboom}, I.~G., et~al. 2006, \mnras, 373, 349

\bibitem[{{Simon} et~al.(2005){Simon}, {Verde}, \&
  {Jimenez}}]{2005PhRvD..71l3001S}
{Simon}, J., {Verde}, L., \& {Jimenez}, R. 2005, \prd, 71, 123001

\bibitem[{{Spergel} et~al.(2007)}]{2007ApJS..170..377S}
{Spergel}, D.~N., et~al. 2007, \apjs, 170, 377

\bibitem[{{Stern} et~al.(2010){Stern}, {Jimenez}, {Verde}, {Kamionkowski}, \&
  {Stanford}}]{2010JCAP...02..008S}
{Stern}, D., {Jimenez}, R., {Verde}, L., {Kamionkowski}, M., \& {Stanford},
  S.~A. 2010, \jcap, 2, 008

\bibitem[{{Stoughton} et~al.(2002)}]{2002AJ....123..485S}
{Stoughton}, C., et~al. 2002, \aj, 123, 485

\bibitem[{{Strauss} et~al.(2002)}]{2002AJ....124.1810S}
{Strauss}, M.~A., et~al. 2002, \aj, 124, 1810

\bibitem[{{Wei}(2010)}]{2010PhLB..687..286W}
{Wei}, H. 2010, Physics Letters B, 687, 286

\bibitem[{{Wu} et~al.(2011){Wu}, {Singh}, {Prugniel}, {Gupta}, \&
  {Koleva}}]{2011A&A...525A..71W}
{Wu}, Y., {Singh}, H.~P., {Prugniel}, P., {Gupta}, R., \& {Koleva}, M. 2011,
  \aap, 525, A71

\bibitem[{{Yi} \& {Zhang}(2007)}]{2007MPLA...22...41Y}
{Yi}, Z.-L., \& {Zhang}, T.-J. 2007, Modern Physics Letters A, 22, 41

\bibitem[{{York} et~al.(2000)}]{2000AJ....120.1579Y}
{York}, D.~G., et~al. 2000, \aj, 120, 1579

\bibitem[{{Zhang} et~al.(2010){Zhang}, {Ma}, \& {Lan}}]{2010AdAst2010E..81Z}
{Zhang}, T.-J., {Ma}, C., \& {Lan}, T. 2010, Advances in Astronomy, 2010,
  184284

\end{thebibliography}

\bibliographystyle{raa}

\end{document}